\input psfig.sty
\documentclass{aa}

\usepackage{graphicx}

\begin{document}



\title{Multicolor Photometry of the Vela Pulsar\thanks{Based
on observations with the NASA/ESA Hubble Space Telescope,
obtained at the Space Telescope Science Institute, which is
operated by AURA, Inc.  under contract No NAS 5-26555}
\thanks{Based on observations collected at the European Southern 
Observatory, La
Silla, Chile}}

\author{R.P. Mignani\inst{1} \and P.A. Caraveo\inst{2}
}

\offprints{rmignani@eso.org}

\institute{ESO, Karl Schwarzchild Str.  2, D-85740 Garching bei
M\"unchen, Germany \and Istituto di Fisica Cosmica del CNR ``G.
Occhialini", Via Bassini 15, I-20133 Milan, Italy}

\date{Received / Accepted }

\titlerunning{Multicolor photometry of the Vela pulsar}

\abstract{Multicolor photometry of the Vela  pulsar (PSR
B0833$-$45), updated by recent  HST/WFPC2 observations obtained in the
$555W$, $675W$  and $814W$ filters, is presented.   The available data
provide the best characterization so far of the pulsar spectral shape,
which is dominated  by a flat power law  continuum with spectral index
$\alpha  = -0.2  \pm$ 0.2,  consistent  with the  extrapolation in  the
optical  domain of  the  power  law component  of  the X-ray  spectrum
detected   by   Chandra.  In   addition,   a  marginally   significant
dip($\approx  3  \sigma$) seems  to  be  present  at about  6500  \AA.
Spectroscopic  observations  with  the  VLT,  now  in  progress,  will
undoubtly  provide  a  much  better  assessment  of  the  reality  and
characteristics   of  this   feature.\keywords{PSRB0833$-$45,  pulsar,
photometry}}

\maketitle

\section{Introduction}  The Vela  pulsar (PSR  B0833$-$45) has  been the
first isolated neutron star to  be identified in the optical after the
Crab   pulsar.   Its   optical   counterpart  ($V=23.6$),   originally
discovered by Lasker (1976), was  found to pulsate at the radio period
(89 ms) by Wallace (1977) with a sharp double-peaked light curve. With
an optical luminosity of about $10^{28}$ erg s$^{-1}$, the Vela pulsar
is the third brightest isolated neutron star after the Crab pulsar and
PSR  B0540$-$69 ,  albeit  with  a much  lower  optical efficiency  as
indicated in  the work of  Goldoni et al.   (1997).  \\ Too old  to be
counted among the young  pulsars ($\approx 1\,000-5\,000$ yrs) and too
young to  belong to the  middle-aged class ($\ge$ 100\,000  yrs), Vela
($\approx$ 10\,000 yrs) appears as a transition object between the two
groups  of neutron stars,  which are  characterized by  very different
emission properties both in the optical and in the X-ray regime. While
in the  young pulsars pure  synchrotron radiation (e.g.   Pacini 1971)
yields  a relatively high  optical emission  efficiency and  power law
spectra  (see e.g.  Hill  et al.   1997 and  Sollerman et  al.  2000),
different emission mechanisms appear at work in older objects. Indeed,
PSR B0656+14 (Pavlov et al.  1997, Koptsevich et al. 2001) and Geminga
(Martin et al.  1998, Mignani  et al. 1998) show more complex spectral
shapes,   coupling  magnetospheric  and   thermal  radiation   with  a
definitely lower emission efficiency.  It is generally agreed that, as
the pulsars  age and  slow down, the  magnetospheric component  of the
optical emission  becomes progressively weakens while  the thermal one
becomes significant  and finally dominates (van  Kerkwijk and Kulkarni
2001).   A similar  evolution  has  been observed  in  the soft  X-ray
regime,  where  young pulsars  exhibit  pure magnetospheric,  strongly
pulsed emission  (Becker \& Tr\"umper  1997) while the  radiation from
middle-aged neutron stars  is well described by a  black body emission
with  shallow modulations.   Vela  is  a unique  example  of a  mostly
thermal soft X-ray emitter (e.g. Pavlov et al.  2001), with just a few
percent  modulation, producing  also totally  pulsed  optical emission
(see,  e.g., Gouiffes  et al.   1998),  albeit with  a low  efficiency
(Goldoni  et al.   1997).  Thus,  characterizing its  optical spectrum
would  be important  to trace  the evolution  in the  emission physics
between the  two groups of  young and middle-aged neutron  stars.\\ So
far, the most  complete optical spectral study of  the Vela pulsar has
been carried out by Nasuti  et al.  (1997) through ground based $UBVR$
bands photometry.  Its multicolor flux  distribution was found  to lie
above any  realistic blackbody curve associated to  the neutron star's
surface emission (\"Ogelman et  al. 1993), which clearly confirmed its
magnetospheric  origin. However, Nasuti  et al.  showed that  the Vela
pulsar  spectrum was  substantially different  from that  of  the Crab
pulsar  and  of PSR  B0540$-$69.  The main  difference  was  due to  a
decrease of  the flux in  the $R$ band,  which was taken by  Nasuti et
al. as  an indication for either  a spectral turnover  at a wavelength
$\ge  6000$ \AA  ~ or  an  absorption feature.  \\ Since  the lack  of
spectral information  at longer wavelenghts  was the main limit  for a
complete  characterization  of  the  Vela  pulsar  spectrum,  we  have
performed new  accurate photometry observations  close to the  $R$ and
$I$ bands  with the WFPC2 on  HST.  A description  of the observations
and of  the data  reduction is  given in Sec.2  while the  results are
discussed in Sec.3.

\section{Observations and Data Reduction}

\begin{table*}[t]
\begin{tabular}{c|c|c|c|c|c|c} \hline
Tel & Instrument & Date & Filter & Wavelength/Width & mag & Flux  \\
      &            &      &        &                  &     &
      ($10^{-29}$ erg cm$^{-2}$s$^{-1}$Hz$^{-1}$) \\ \hline
HST & WFPC2 & 19.03.2000 & 814W & 7995 \AA (1292\AA) & 24.40 (0.1) & 1.69
(0.16) \\
NTT & EMMI  & 30.01.1995 & R    & 6410 \AA (1540 \AA)& 23.93 (0.2) & 0.98
(0.20) \\
HST & WFPC2 & 15.03.2000 & 675W & 6717 \AA (1536 \AA)& 24.30 (0.1) & 1.38
(0.13)\\
NTT & EMMI  & 30.01.1995 & V    & 5426 \AA (1044 \AA)& 23.65 (0.1) & 1.84
(0.16)\\
HST & WFPC2 & (*)        & 555W & 5442 \AA (1044 \AA)& 23.64 (0.05)& 1.82
(0.08) \\
NTT & EMMI  & 30.01.1995 & B    & 4223 \AA (941 \AA) & 23.89 (0.15)& 1.88
(0.26) \\
NTT & EMMI  & 30.01.1995 & U    & 3542 \AA (542 \AA) & 23.38
(0.15)& 1.61 (0.22) \\ \hline
\end{tabular}

\vspace{0.5cm} (*)
30.06.1997,2.01.1998,30.06.1999,15.01.2000,5.07.2000
\vspace{0.5cm}

\caption{Summary of the available multicolor  photometry  of the
Vela  pulsar.  The telescopes  and detectors  used, together  with the
observing   epochs,   are  listed   in   the   first  three   columns,
respectively. The filter names are reported in the fourth column while
their  pivot wavelength  and  width  are in  the  fifth. The  observed
magnitudes, computed  wrt to the Johnson and  HST photometric systems,
and attached errors are listed  in column six.  Column seven gives the
corresponding  fluxes  and  errors  at  the  pivot  wavelengths  after
dereddening for an  $A_V = 0.4$ (Manchester et  al.  1978). The $555W$
magnitude is the average  of independent measurements obtained at five
different epochs.}
\end{table*}

The Vela  pulsar was observed with  the WFPC2 aboard the  HST on March
15th  and 19th  2000.   The observations  were  performed through  the
filters $675W$ ($\lambda = 6717$  \AA,$\Delta \lambda = 1536$ \AA) and
$814W$ ($\lambda = 7995$ \AA,  $\Delta \lambda = 1292$ \AA). While the
observation  in the  $675W$ filter  was aimed  at confirming  the flux
value of Nasuti et al. in  the $R$ band, the observation in the $814W$
one was taken to disentangle the  presence of an absorption dip from a
real spectral turnover.  In addition,  WFPC2 images of the Vela pulsar
were  repeatedly  obtained in  the  $555W$  filter  ($\lambda =  5442$
\AA,$\Delta \lambda  = 1044$ \AA) as  a part of a  set of observations
aimed  at the  measurement of  the  neutron star's  parallax.  In  all
cases, the exposure time was set  to 2600 s, corresponding to a single
orbit of the spacecraft.  The  pulsar was always located at the center
of the Planetary Camera ($PC$) chip of the WFPC2, with a corresponding
pixel size of 45.5  mas and a field of view of  $35 \times 35$ arcsec.
All the observations were split  into shorter exposures to allow for a
better cosmic rays filtering.   After the standard pipeline processing
to remove instrumental effects, single exposures were combined using a
median filter algorithm.  Flux  calibration was computed following the
standard WFPC2  recipe (e.g. Holtzmann  et al. 1995).  The  pulsar was
clearly detected  in all passbands  and its magnitude was  computed by
standard  aperture  photometry  using  the  same  optimized  areas  to
estimate the counts both from  the source and from the sky background.
The magnitudes in  the three passbands $814W$, $675W$  and $555W$ were
thus computed  to be  24.40 ($\pm$ 0.1),  24.30 ($\pm$ 0.1)  and 23.64
($\pm$ 0.05),  respectively, with the photometric error  in the $555W$
filter being averaged over the five epochs.

\begin{figure}
\centerline{\hbox{\psfig{figure=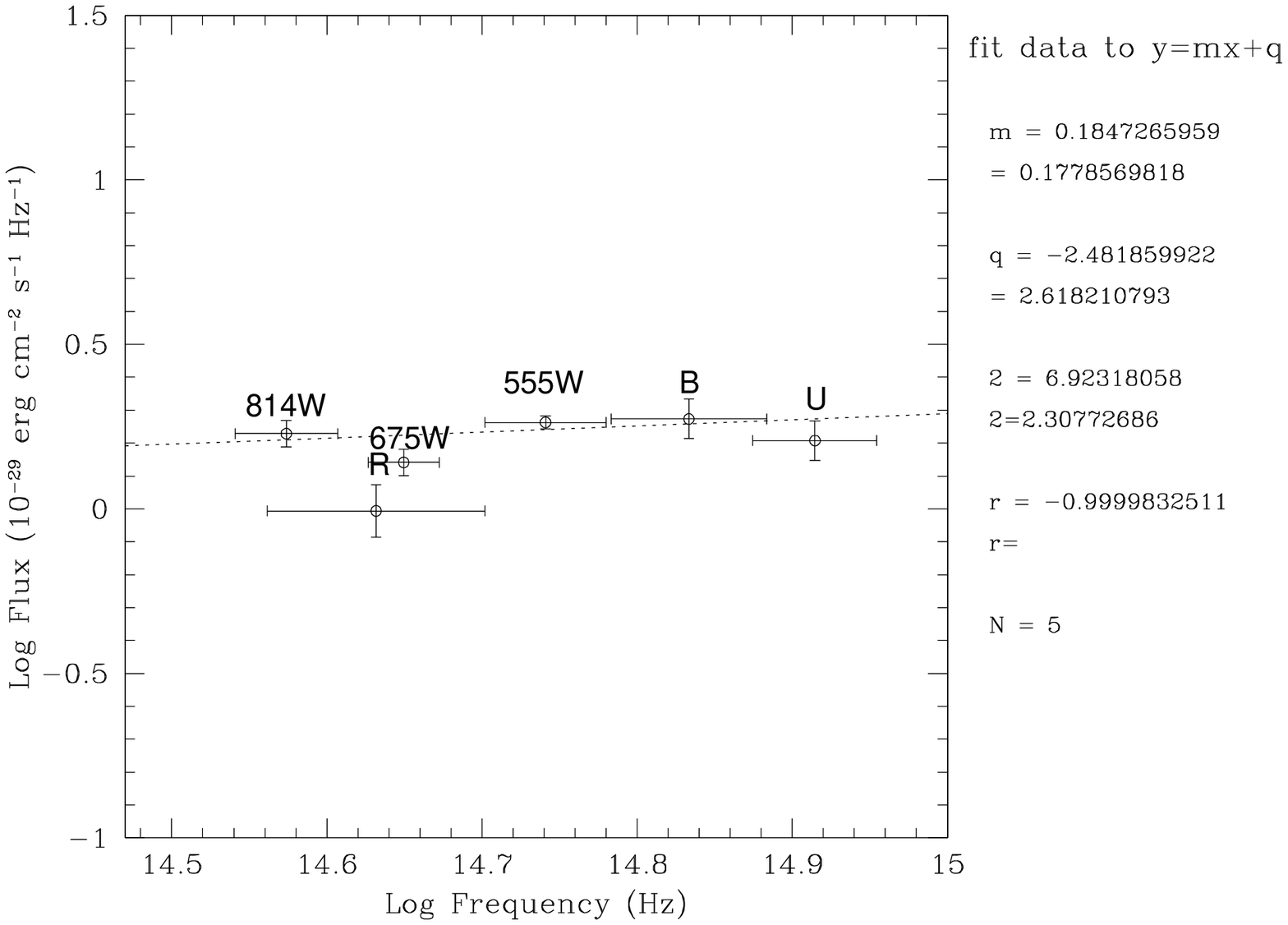,angle=0,width=9cm,clip=} }}
\caption{Multicolor fluxes of PSR0833$-$45 computed from the
magnitudes listed in Table 1.  The $UBVR$ fluxes have been obtained by
the NTT (Nasuti et al.  1997),  while the ones labelled with the three
digits  ($814W$,  $675W$  and  $555W$)  have been  obtained  with  the
HST/WFPC2. The $V$  point virtually coincides with the  $555W$ one and
is not plotted  for clarity. The dashed line  represents the power law
($\alpha = -0.2 \pm 0.2$) bestfitting the available data set  with the
exception of  the $R$ point, which  appears to deviate  from the trend
and  suggests the possible presence of a dip around 6\,500 \AA.  }
\label{Fig. 1}
\end{figure}

\section{Results}
The complete multicolor photometry  of the Vela pulsar, including both
the  newly   acquired  HST  values  and  the   previous  ground  based
measurements  of  Nasuti  et  al.   (1997),  is  summarized  in  Table
1. Magnitudes  have  been corrected  for  the interstellar  extinction
using  as  a reference  $A_V  = 0.4$,  as  reported  in Manchester  et
al. (1978).   The corresponding flux  values are plotted in  Figure 1,
where the  HST data  are seen to  improve significantly on  the ground
based colours of Nasuti et  al.   In particular, the spectrum shown in
Figure 1, with the first flux measurement in the $I$-band, provides no
indication for  the spectral turnover  suggested by Nasuti et  al.  on
the basis  of their limited  data set.   The spectral  distribution is
compatible with a power law spectrum ($F_\nu \propto \nu^{- \alpha}$),
as  expected in  the  case of  pure  magnetospheric radiation  (Pacini
1971). A  fit to the points shown  in Figure 1, with  the exception of
the clearly deviant  $R$ point, gives a spectral  index $\alpha = -0.2
\pm  0.2$,   which is  comparable  with the  one  of  the Crab  pulsar
($\alpha = 0.1 \pm  0.01$) obtained from spectroscopic observations by
Nasuti  et al.   (1996) and  more recently  confirmed by  Sollerman et
al. (2000).   In particular, is  interesting to note that the slope of
the Vela spectral distribution is consistent with the extrapolation of
the power law, magnetospheric component measured by CHANDRA (Pavlov et
al.  2001)  for E $\ge$ 1.8  keV. This points towards  a common origin
for both the optical and  the X-ray emissions. \\ Although the optical
spectrum of  Vela is dominated by  a flat continuum,  a deviation from
the trend can be seen around 6\,500
\AA, where the $675W$ point seems to follow the trend for a flux
decrease suggested by the ground  based, less accurate, $R$ point.  On
the other hand, the $814W$ measurement is higher than both the $R$ and
$675W$  ones. Applying  different  reddening corrections  to our  data
using, e.g.   the $N_H$ obtained from  the spectral fits  to the ROSAT
(\"Ogelman 1993; Page  et al.  1996) or CHANDRA  (Pavlov et al.  2001)
data  does  not  alter  substantially  our result.   Thus,  the  newly
acquired  colours suggest the  presence of  a dip  in the  spectrum at
$\approx$ 6\,500 \AA.  Within the uncertainty of our photometry and of
the  flux  calibration, we  can  estimate  its  significance at  about
$\approx 3 \sigma$.  However, given the marginal statistical weight of
this  detection, admittedly  affected  by the  lower  accuracy of  the
ground-based $R$  point, any interpretation of the  physical origin of
the dip would be premature.   \\ The spectral distribution of the Vela
pulsar can be compared with  those of the other isolated neutron stars
with  multicolor photometry.   This is  done  in Figure  2, where  the
objects' fluxes have been scaled  to fit into the same frame. Although
limited to  five objects  (the Crab pulsar,  PSR B0540$-$69,  the Vela
pulsar, PSR  B0656+14 and Geminga),  this sample covers more  than two
decades in  pulsar age and  ten magnitudes in  flux going all  the way
from the bright, 1\,000 yrs  old, Crab to the 25.5 magnitude, 500\,000
yrs old, Geminga (see, e.g.,  Caraveo 2000 for a recent review).  From
the  youngest  to  the  oldest,  the flux  distributions  increase  in
complexity going from monotonic,  power-law spectra to composite ones,
where the thermal  emission from the hot neutron  star surface adds to
the magnetospheric  one (Mignani  et al.  1998;  Pavlov et  al. 1997).
 Such an evolutionary interpretation is proposed also by Koptsevich et
al.   (2001) who,  on the  basis of  a similar  comparative multicolor
study, aimed at putting in  context their new results on PSR B0656+14,
remark  the  non  monotonic   behaviour  seen  both  for  Geminga  and
PSR0656+14, which "might indicate  the presence of unresolved spectral
features".   While  waiting  for  spectroscopic observations  of  more
isolated  neutron  stars,  we  can  safely  state  that  the  spectral
distributions  measured for  middle-aged isolated  neutron  stars look
different  from that  of  the  very young  ones,  underlining the  key
position  of the  Vela  pulsar . \\Both  age-wise  and flux-wise  Vela
appears  to  occupy an  "in  between"  position.  Its flat,  power-law
spectrum, coupled  with its 100\%  pulsation, makes it similar  to the
younger  objects, while  its low  efficiency brings  it nearer  to the
older ones.  

\begin{figure}
\centerline{\hbox{\psfig{figure=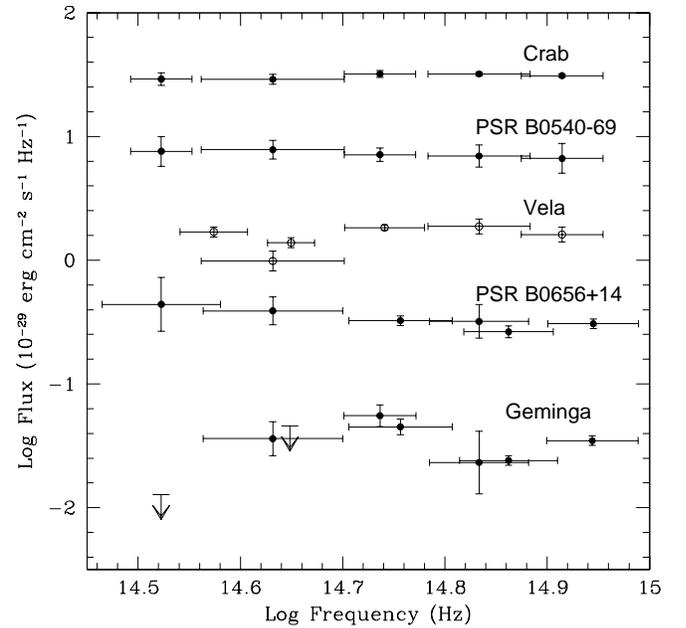,angle=0,width=8.8cm,clip=}}}
\caption{The  $I$ to $U$ colors of  the Vela pulsar (open
circles)  are  compared  to  those  of the  other  optically  emitting
isolated neutron stars (filled circles).  From top to bottom: the Crab
pulsar  and PSR  B0540$-$69  (Nasuti  et al.  1997),  PSR 0656+14  and
Geminga (Koptsevich et al. 2001;  Mignani et al.  1998).  To allow for
a  better representation on  the same  frame, the  fluxes of  the Crab
pulsar and Geminga have been re-scaled. } \label{Fig.  2}
\end{figure}

\section{Conclusions}

We have reported on new colours of  the Vela pulsar in the red part of
the spectrum, obtained with  the HST/WFPC2.  These observations, which
also provide the  first detection of the pulsar in  the $I$ band, have
been used to improve  and complement the previous ground-based results
of  Nasuti et  al.   (1997).  The  pulsar  spectral distribution,  now
determined with  unprecedented detail, is clearly dominated  by a flat
power  law  continuum ($\alpha  =  -0.2  \pm  0.2$) with  no  spectral
turnover.   In addition, a marginal  indication for a dip is found at
$\approx$  6\,500 \AA.   The first  spectroscopic observations  of the
Vela pulsar, now  in progress with the VLT, are  needed to confirm the
reality of such a dip and to address its origin.

\begin{acknowledgements}
  We wish to thank the anonymous referee for his/her comments, which helped 
to improve the quality of the paper.
\end{acknowledgements}

\end{document}